\documentclass[11pt]{revtex4-2}
\usepackage{relsize}
\usepackage{hyperref}
\usepackage{amsmath,amsfonts,amssymb}
\hypersetup{dvips,dvipdfm,colorlinks=true,urlcolor=magenta,filecolor=magenta,linktoc=page,citecolor=red,linkcolor=blue,bookmarks=true}
\usepackage{graphicx,epstopdf}

\begin{document}
	\title{Is warm inflation quasi-stable?}
\author{Akash Bose$^1$\footnote {bose.akash13@gmail.com}}
\author{Subenoy Chakraborty$^1$\footnote {schakraborty.math@gmail.com (Corresponding Author)}}
\affiliation{$^1$Department of Mathematics, Jadavpur University, Kolkata-$700032$, West Bengal, India}
	\begin{abstract}
	The present work deals with a non-equilibrium thermodynamics that is associated with the scenario of warm inflation. The premise is that an adiabatic radiation production process holds exactly i.e. the radiation dilution is exactly counterbalanced by a dissipation term. Under this hypothesis, it is found that radiation particle number, temperature, radiation energy density and pressure are all conserved -- a contradiction to the very nature of the warm inflation dynamics. However, such exact adiabatic radiation production process never happens in any realistic analysis of warm inflation. In the slow roll approximation this holds at best at the zeroth order. Finally it is shown that a variable cosmological constant may accommodate the quasi-stable process in warm inflation with non-equilibrium thermodynamic description.
\end{abstract}
\maketitle
This is the 41st year since the idea of cosmological inflation was introduced by A. H. Guth \cite{Guth:1980zm}. Today inflation is widely accepted leading theory of the early universe due to its success in solving many long-standing puzzles of the hot big bang model namely the horizon, flatness and monopole problems \cite{Guth:1980zm,Starobinsky:1980te}. Moreover, due to the very high energy scale (200 Gev - $10^{12}$ Tev) \cite{Martin:2013tda} in inflationary era, the inherent quantum effects can explain the origin of seeds for large-scale structure (LSS) formation as well as the fluctuations generating the cosmic microwave background (CMB) anisotropies \cite{Mukhanov:1981xt}. So inflationary paradigm can be considered as one of the corner stones of modern cosmology.

For the last few years we have a large number of precision data from various observational setup namely cosmic microwave background (CMB) \cite{Aghanim:2015xee} to large scale structures \cite{Scolnic:2017caz}, including Baryon Acoustic oscillation data \cite{Dawson:2012va}, both strong and weak lensing \cite{Abbott:2017wau}, galaxy cluster number counts \cite{Ade:2015gva} and so on up to gravitational waves detection \cite{Abbott:2016blz}. These observational data not only predict the nature and the evolution of the Universe but also indicate the mechanisms operating at the very early times. Thus inflationary scenario is a very predictive scenario in the sense that it was developed long before the above sophisticated data were available.

Usually, an important physical candidate for generating inflationary phase for the very early evolution is a single scalar field model having logarithm of the potential moves much slower than the kinetic part to have rapid exponential expansion. An important issue for the inflationary scenario is to formulate a physically realistic mechanism by which the early de sitter type accelerated expansion can smoothly enter into radiation epoch of the standard big bang model. Depending on the dynamics of the inflaton field, the notion of inflation can have two distinguished classifications namely cold inflation and warm inflation.

In cold inflation (CI), the end of inflation and the beginning of the hot big bang model are correlated through the quantum fluctuations due to the scalar field oscillations. However, due to lack of observational evidences, the preheating and reheating eras are purely adhoc in nature. Further, in this set up, in addition to the adiabatic initial condition on the CMB, the temperature due to reheating has to be larger than the big bang nucleosynthesis (BBN) scales \cite{Martin:2010kz}. Also the interaction of the inflaton with other field degrees of freedom is so weak that it is not possible to counter balance the dilution of (preexisting or newly formed) radiation. As a result, the universe goes over to a super freezed era and quantum fluctuations of the inflaton field originate density perturbations. On the other hand, warm inflation (WI) scenario can be described from field theoretic view point if the radiation field is in thermal equilibrium. Here the interaction between the inflaton and other fields is so significant that a quasi-stationary thermalized radiation bath is formed. The thermal fluctuations of this radiation bath is the primary source for the density fluctuations which is transported to the inflaton field as adiabatic curvature perturbation \cite{Taylor:2000ze,Hall:2003zp}. Due to this dissipative effect non-trivial dynamics in WI has significant effect on observational quantities such as tensor-to scalar ratio ($r$), the spectral index ($n_s$) and the non-Gaussianity parameter ($f_{NL}$) \cite{Bartrum:2013fia} and hence WI can well be tested observationally. Further in WI the radiation production takes place during the inflationary expansion (driven by inflaton) so there is no need of any reheating era. This is possible for the friction term in the evolution equation for the inflaton field and consequently, the scalar field dissipates into a thermal bath with other fields.

Moreover, the strong coupling between the inflaton and other fields has produced a significant radiation production rate, preserving the expected flatness of the inflaton potential and consequently the radiation so produced prevents super cooling (in CI) of the Universe, so that there is a possible smooth transition to the radiation dominated phase of standard big bang model (without any pre/reheating phase) after the end of the inflationary regime. Another issue in favor of WI is the initial fluctuations responsible for the LSS formation.

 	In the present work, WI is considered in the background of spatially flat FLRW model with matter component in the form of inflaton (a scalar field with self-interacting potential) interacting with radiation. The Friedman equations can be written as 
 \begin{equation}\label{eq1}
 3H^2=\left(\rho_\phi+\rho_r\right),~~~~~ 2\dot{H}=-(\rho_\phi+p_\phi)-(\rho_r+p_r)
 \end{equation}
 (choosing $c=1,\kappa=8\pi G=1$). Here $H=\dfrac{\dot{a}}{a}$ represents Hubble parameter and the quantity $`a$' denotes scale factor; `overdot' represents differentiation with respect to time.  $\rho_\phi=\dfrac{1}{2}\dot{\phi}^2+V(\phi)$ is the energy density of the homogeneous scalar field $\phi=\phi(t)$, $V(\phi)$ is the effective potential and $\rho_r$ is the energy density of the radiation field. The total energy density $\rho=\rho_\phi+\rho_r$ and the continuity equation for the total energy density satisfies the standard relation $\dot{\rho}+3H(\rho+p)=0$. Also $p_\phi$ and $p_r$ are the pressure of the scalar field and radiation field respectively, given by $p_\phi=\dfrac{1}{2}\dot{\phi}^2-V(\phi)$ and $p_r=\dfrac{1}{3}\rho_r$.
 
 During the scenario of warm inflation, the scalar and radiation component interact. Due to this interaction, energy is transferred from the scalar field to radiation fluid. This process is described through the conservation equations:
 \begin{equation}\label{eq2}
 \dot{\rho}_\phi+3H(\rho_\phi+p_\phi)=-\Gamma\dot{\phi}^2,~~~~~\dot{\rho}_r+3H(\rho_r+p_r)=\Gamma\dot{\phi}^2
 \end{equation} 
 or equivalently
 \begin{equation}\label{eq3}
 \ddot{\phi}+(3H+\Gamma)\dot{\phi}+\frac{\partial V}{\partial\phi}=0\mbox{~~i.e.~~}\ddot{\phi}+3H(1+Q)\dot{\phi}+\frac{\partial V}{\partial\phi}=0 
 \end{equation} 
 
 Here $\Gamma$ is termed as the dissipation coefficient and $Q=\dfrac{\Gamma}{3H} $ is the ratio of the radiation production to expansion rate. From second law of thermodynamics $\Gamma$ should be positive, indicating an energy flow from inflaton to the radiation fluid.
 
 	Now, in warm inflationary era, the energy density of the scalar field predominates over the energy density of the radiation field i.e. $\rho_\phi>\rho_r$. Also in slow-roll approximation i.e. $\dot{\phi}^2\ll V(\phi)$ and $\ddot{\phi}\ll(3H+\Gamma)\dot{\phi}$, the Friedmann equation (\ref{eq1}) approximates as
 \begin{equation}
 3H^2\simeq\rho_\phi\simeq V(\phi),\label{eq4}
 \end{equation}
 while the scalar field evolution equation (\ref{eq3}) takes the form
 \begin{equation}\label{eq6}
 \dot{\phi}\simeq-\frac{\frac{\partial V}{\partial \phi}}{3H(1+Q)},
 \end{equation}
 
 Depending on the magnitude of $Q$ one has two regions namely weak dissipative regime ($Q\ll$ 1) and strong dissipative regime ($Q\gg1$). Further, assuming quasi-stable nature of the radiation production (during inflationary epoch) \cite{Berera:1995ie,Berera:1995wh} one should have $\dot{\rho}_r\ll4H\rho_r,\Gamma\dot{\phi}^2$ and hence the evolution equation (\ref{eq3}) approximates as (using (\ref{eq6}))
 \begin{equation}\label{eq6a}
 \rho_r=c_\gamma T^4\simeq\frac{\Gamma\dot{\phi}^2}{4H}=\frac{Q}{4(1+Q)^2}\frac{\left(\frac{\partial V}{\partial \phi}\right)^2}{V},
 \end{equation}
 where $c_\gamma=\dfrac{\pi^2g_\ast}{30}$ is a constant with $g_\ast$ indicating the number of relativistic degrees of freedom \cite{Hall:2003zp,Berera:1995ie}. So the temperature of the thermal bath has the expression
 \begin{equation}
 T=\left[\frac{Q}{4c_\gamma (1+Q)^4}\frac{\left(\frac{\partial V}{\partial \phi}\right)^2}{V}\right]^\frac{1}{4}.
 \end{equation}
 
 Now, the number of e-folding $N$, a measure of the inflationary expansion of the Universe, can be expressed as (using (\ref{eq4}) and (\ref{eq6})) 
 \begin{equation}
 N=\int\limits_{t}^{t_e} H dt'=\int\limits_\phi^{\phi_e}\frac{V(\phi')(1+Q)}{\frac{\partial V}{\partial \phi'}}d\phi'.
 \end{equation}
 
Moreover, in WI due to the presence of the radiation field, the source of the density fluctuations are due to thermal fluctuation \cite{Berera:1995ie,Hall:2003zp} so that the fluctuations of the scalar field are dominated by thermal, rather than quantum \cite{Berera:1995ie,Berera:1995wh} in nature. As in WI the mixture of the scalar field and radiation are produced at the perturbative levels so the curvature and entropy perturbations coexist. But it has been shown \cite{Hall:2003zp} that during WI the entropy perturbations decay while the curvature (as adiabatic modes) perturbation survives \cite{Berera:1995ie,Berera:1995wh}, so the power spectrum of the curvature perturbation (in slow roll approximation) can be written as (assuming $\Gamma=\Gamma(\phi)$) \cite{Herrera:2018cgi}
\begin{equation}
	P_s\simeq\frac{H^3 T}{\phi^2}\sqrt{1+Q}.
\end{equation}

Also the scalar spectral index defined as $n_s=\dfrac{d\ln P_s}{d\ln k}$ has the expression \cite{Moss:2008yb}
\begin{equation}
	n_s=1-\frac{(9Q+17)}{4(1+Q)^2}\epsilon-\frac{(9Q+1)}{4(1+Q)^2}\beta+\frac{3}{2}\frac{1}{(1+Q)}\eta,
\end{equation}
with \begin{equation}
	\epsilon=\dfrac{1}{2}\dfrac{\left(\frac{\partial V}{\partial \phi}\right)^2}{V^2},\qquad \eta=\dfrac{V_{,\phi\phi}}{V},\qquad \beta=\dfrac{V_{,\phi}\Gamma_{,\phi}}{V\Gamma},
	\end{equation} the slow roll parameters.

As there is no coupling between the tensor perturbation with thermal background so the tensor modes have an equivalent amplitude as in cold inflation (i.e$.$ the tensor spectrum $P_T=8H^2$) and hence the tensor to scalar ratio $r$ rakes the form
\begin{equation}
	r=\frac{P_T}{P_s}=\frac{16\epsilon}{(1+Q)^\frac{5}{2}}\frac{H}{T}.
\end{equation}
 
Usually, in WI the reheating period can be eliminated considering decay of the inflaton particle number and radiation particles will be gradually produced during inflationary era. As a result, there will have a smooth transition from the inflationary phase into the radiation dominated phase. So one naturally assumes that the above created (radiation) particles give rise to a thermal gas of radiation. The damping term in the evolution equation of the inflaton field is responsible for the creation of radiation
particle. Hence in the prescription of non-equilibrium thermodynamics the above WI model can be considered as an open thermodynamical system with non-conservation of individual 
fluid particle number
 i.e$.$ $N^\alpha_{I_;{_\alpha}} \neq0$
\begin{equation}\label{eq5}
N^\alpha_{I_;{_\alpha}}\equiv \dot{n}_I+\Theta n_I=n_I\Pi_I
\end{equation}

Here $\Theta=u^\alpha_{I_;{_\alpha}}$ is the fluid expansion, $u^\alpha_I$ is the fluid four velocity, $N^\alpha_I=n_Iu^\alpha_I$ is the particle flow vector, $n_I$ is the  particle number density, $\dot{n}_I=n_{I_;{_\alpha}}u^\alpha_I$, $\Pi_I$ represents the rate of change of the number of particles in a comoving volume $a^3$ and suffix $I=(\phi,r)$ stands for the above two fluids. Here $\Pi_I$ can take both signs; $\Pi_I>0$ implies creation of particles while $\Pi_I<0$ indicates annihilation of particles.

It should be noted that any nonzero interaction term $\Gamma$ may act as an effective bulk viscous pressure so that the matter conservation equations (\ref{eq2}) can be written as
\begin{equation}\label{eq15}
\dot{\rho}_I+3H(\rho_I+p_I+p_I^c)=0,
\end{equation}
with $p^c_\phi=Q\dot{\phi}^2>0$ and $p^c_r=-Q\dot{\phi}^2<0$ as the effective bulk viscous pressure.
%

Suppose in a closed thermodynamical system there are $N$ no. of particles having internal energy $E$, the first law of thermodynamics which is essentially the conservation of internal energy, can have the differential form
\begin{equation}
dE=dQ-pdV
\end{equation}

where $dQ$ is the amount of heat received by the system in a comoving volume $V$in time $dt$. Using Clausius relation the above conservation equation can be written in the form of Gibbs equation as
\begin{equation}
Tds=dq=d\left(\frac{\rho}{n}\right)+pd\left(\frac{1}{n}\right)
\end{equation}

where $\rho=\dfrac{E}{V}$ is the energy density, $n=\dfrac{N}{V}$ is the particle number density, $dq=\dfrac{dQ}{N}$ is the heat per unit particle, $s$ is the entropy per particle and $T$ is the temperature of the system. It is interesting to note that Gibbs equation does not depend on particle number conservation (or non-conservation)

So in the present interacting two fluid system (where particle numbers are not conserved) Gibbs equation takes the form: \cite{Chakraborty:2014ora}
\begin{equation}\label{eq8}
T_Ids_I=d\left(\frac{\rho_I}{n_I}\right)+p_Id\left(\frac{1}{n_I}\right)
\end{equation}
for the $I$-th fluid and is the starting point for studying non-equilibrium thermodynamics. Using the individual conservation equations for the fluid system and the particle numbers, the above Gibbs equation has the explicit form as
\begin{equation}\label{eq9}
n_IT_I\dot{s}_I=-\Theta p^c_I-\Pi_I(\rho_I+p_I)
\end{equation}

Assuming the thermodynamical system to be isentropic (adiabatic) in nature, the entropy per particle remains constant and hence from equation (\ref{eq9}) the bulk viscous pressure depends linearly on the particle creation rate as
\begin{equation}\label{eq10}
p^c_I=-\frac{\Pi_I}{\Theta}(\rho_I+p_I)
\end{equation}

So a dissipative fluid may be considered as equivalent to a perfect fluid with non-constant particle number. However, in adiabatic process there is entropy variation both due to particle number variation as well as due to the enlargement of the phase space of the system.

In the second order formulation of the non-equilibrium thermodynamics due to Israel and Stewart, the entropy flow vector of the $I$-th fluid is given by
\begin{equation}
S^\alpha_I=s_IN^\alpha_I-\frac{\tau {p^c_I} ^2}{2\zeta_IT_I}u^\alpha_I
\end{equation}

where $\tau$ is the relaxation time and $\zeta_I$ is the coefficient of bulk viscosity of the $I$-th fluid. Now using non-conservation of particle number (i.e., equation (\ref{eq5}) and the Gibbs equation (\ref{eq8}) (or equation (\ref{eq9}))) one gets
\begin{equation}
T_IS^\alpha_{I_;{_\alpha}}=-n_I\mu_I\Pi_I-p^c_I\left[\Theta+\frac{\tau\dot{p}^c_I}{\zeta_I}+\frac{1}{2}p^c_IT_I\left(\frac{\tau}{\zeta_IT}u^\alpha_I\right)_{;\alpha}\right]
\end{equation}

where $\mu_I=\dfrac{\rho_I+p_I}{n_I}-T_Is_I$ is the chemical potential. So the choice of the generalized ansatz 
\begin{equation}
\Theta+\frac{\tau\dot{p}^c_I}{\zeta_I}+\frac{1}{2}p^c_IT_I\left(\frac{\tau}{\zeta_IT}u^\alpha_I\right)_{;\alpha}+\frac{\mu_In_I\Pi_I}{p^c_I}=-\frac{p^c_I}{\zeta_I}
\end{equation}
guarantees the second law of thermodynamics i.e. 
$$S^\alpha_{I_;{_\alpha}}=\frac{{p^c_I}^2}{\zeta_IT_I}\geqslant0$$

As a result, the effective viscous pressure $p^c_I$ (of the $I$-th fluid) has the non-linear evolution equation as
\begin{equation}
T_I\frac{d}{dt}\left({p^c_I}^2\right)+2{p^c_I}^2+\zeta_IT{p^c_I}^2\left(\frac{\tau}{\zeta_IT}u^\alpha_I\right)_{;\alpha}+2\zeta_Ip^c_I\Theta=-2\zeta\mu_In_I\zeta_I\Pi_I
\end{equation}

One may note that the chemical potential is responsible for the above non-linearity and hence it may act as an effective symmetry-breaking parameter in relativistic field theories.

The present second order theory is also known as causal theory (non-vanishing relaxation time) compare to the first order Eckart theory as non-causal (i.e$.$ vanishing relaxation time). Physically, the difference between these two theories can be distinguished as follows: Here the effective bulk viscous pressure $p^c_I$ appears due to particle creation rate $\Pi_I$. In causal theory, if $\Pi_I$ disappears (i.e. switch off) then $p^c_I$ decays to zero over the relaxation time $\tau$ while in non-causal theory this will happen instantaneously.

Now choosing the particle number density $n$ and the temperature $T$ as the basic thermodynamical variables the equation of state in general form can be written as
\begin{equation}
\rho=\rho(n,T)\mbox{~and~}p=p(n,T)
\end{equation}

Using the above equation of state and the conservation equations (for fluids and number density) in the general thermodynamical relation
\begin{equation}
\frac{\partial\rho}{\partial n}=\frac{\rho+p}{n}-\frac{T}{n}\frac{\partial p}{\partial T}
\end{equation}
one obtains the evolution equation for the temperature as
\begin{equation}
\frac{\dot{T}_I}{T_I}=-\Theta\left[\frac{p^c_I}{T_I\left(\frac{\partial\rho_I}{\partial T_I}\right)}+\frac{\frac{\partial p_I}{\partial T_I}}{\frac{\partial\rho_I}{\partial T_I}}\right]+\Pi_I\left[\frac{\frac{\partial p_I}{\partial T_I}}{\frac{\partial\rho_I}{\partial T_I}}-\frac{\left(\rho_I+p_I\right)}{T_I\left(\frac{\partial\rho_I}{\partial T_I}\right)}\right]
\end{equation}
which on simplification (using equation (\ref{eq9})) gives
\begin{equation}
\frac{\dot{T}_I}{T_I}=-\left(\Theta-\Pi_I\right)\frac{\frac{\partial p_I}{\partial T_I}}{\frac{\partial\rho_I}{\partial T_I}}+\frac{n_I\dot{s}_I}{\frac{\partial\rho_I}{\partial T_I}}
\end{equation}

Further, assuming the thermodynamical system to be adiabatic in nature then using equation (\ref{eq10}) gets more simplified as
\begin{equation}\label{eq19}
\frac{\dot{T}_I}{T_I}=-\left(\Theta-\Pi_I\right)\frac{\partial p_I}{\partial\rho_I}
\end{equation} 
while the other thermodynamical variables evolve as
\begin{eqnarray}\label{eq20}
\begin{split}
\dot{\rho}_I&=&-(\Theta-\Pi_I)(\rho_I+p_I),~~\\
\dot{p}_I&=&-c_s^2(\Theta-\Pi_I)(\rho_I+p_I),\\
\mbox{and~~}\frac{\dot{n}_I}{n_I}&=&-(\Theta-\Pi_I)~~~~~~~~~~~~~~~
\end{split}
\end{eqnarray}
where $c_s^2=\left(\dfrac{\partial p_I}{\partial\rho_I}\right)_{\text{adia}}$, is the (square) adiabatic sound speed.

Now one can write equation (\ref{eq10}) explicitly as
\begin{eqnarray}\label{eq21}
p^c_\phi=-\frac{\Pi_\Phi}{\Theta}(\rho_\phi+p_\phi)\mbox{~~~~~~~~~and~~~~~~~~~~~~}p^c_r=-\frac{\Pi_r}{\Theta}(\rho_r+p_r)\nonumber\\
\implies Q\dot{\phi}^2\Theta=-\Pi_\Phi\dot{\phi}^2,~~~~~~~~~~~~~~~~~~~~~~\implies-Q\dot{\phi}^2\Theta=-\Pi_r\cdot\frac{4}{3}\rho_r\nonumber\\
\implies\Pi_\Phi=-3HQ=-\Gamma<0,~~~~~~~~~~~~~\implies\Pi_r=\frac{3}{4}\frac{\Gamma\dot{\phi}^2}{\rho_r}>0~~~~~
\end{eqnarray}

This implies inflaton particles annihilate and radiation particles create in accordance with WI scenario. If the radiation production during inflation is assumed to be quasi-stable in nature then $\dot{\rho}_r\ll H\rho_r,\Gamma\dot{\phi}^2$, So the radiation energy conservation relation gives $\rho_r\approx\dfrac{\Gamma\dot{\phi}^2}{4H}$ and one can simplify equation (\ref{eq20}) and equation (\ref{eq21}) as
\begin{eqnarray}
\Pi_r\approx3H,~~~~ \Pi_\Phi=-3HQ\\
\frac{\dot{n}_r}{n_r}=-(\Theta-\Pi_r)\approx0~~~
\end{eqnarray}

Hence radiation particle number is conserved. Further from equations (\ref{eq19}) and (\ref{eq20}), all the physical quantities (like temperature, energy density, pressure) of radiation particle is conserved. This is in contradiction to the WI scenario where the Universe should gradually dominated by radiation. Also particle number conservation  is not in favor of non-equilibrium thermodynamics. One can overcome this issue either assuming non-adiabatic nature of the thermodynamical system (i.e., equation (\ref{eq10}) does not hold) or the radiation production process should not be quasi-stable in nature (i.e., $\rho_r\not\approx\dfrac{\Gamma\dot{\phi}^2}{4H}$ and hence $\Pi_r\not\approx3H$). So, one may conclude that adiabatic thermodynamical analysis of WI does not provide quasi-stable radiation process.

However, such exact adiabatic radiation production process never happens in any realistic analysis in warm inflation. In fact, in the radiation evolution equation (\ref{eq2}) the above conclusion would imply that the dissipation term, which acts as a source term in that equation, would exactly counterbalance the dilution term due to the expansion. But this can only hold approximately, at best in a zeroth-order in the slow-roll approximation. Also the approximated relation in equation (\ref{eq6a}) receives corrections from slow-roll terms and it is, thus, not an exact expression (for details see \cite{Das:2020lut}). In these works it has been shown explicitly that at near the end of warm inflation, just before the radiation energy density overtakes the inflaton energy density, all quantities change quickly, which is just a consequence of the slow roll approximation no longer holding.

Now it is interesting to see the consequence of quasi-stable scenario in the modified gravity theories.

Any modified gravity theory can be interpreted as the Einstein gravity with two fluid system of which one is the usual fluid and the other one is hypothetical effective fluid whose energy density and pressure is given by the extra terms in the Friedmann equation of the corresponding modified gravity theory. In the context of WI, the effective fluid is chosen as the inflaton fluid while the usual fluid is considered as the radiation fluid. Depending on the nature of these two fluids, the modified gravity can be classified into two types.

(i) The fluids are non-interacting, i.e., the continuity equation of the corresponding modified gravity theory is given by $\dot{\rho}+3H(p+\rho)=0$. (For example $f(R)$ gravity, $f(T)$ gravity, $f(R,T)$ gravity and so on). Since there is no interaction, there is no energy transfer between these two fluids and consequently non-equilibrium thermodynamics cannot come into scenario. Hence there is no possibility of production of radiation particle. So warm inflation is not possible for those types of modified gravity theories.

(ii) The fluids are interacting, i.e, there is an extra term in the right hand side of the continuity equation. (For example Einstein-Cartan-Kibble-Sciama (ECKS) gravity theory, fractal gravity and so on). One can termed this extra term as interaction term $\mathcal{I}$ and consequently the evolution equation of the individual fluid takes the form of equation (\ref{eq2}). Now proceeding as above, one can finally obtain the same conclusion $\dot{n}_r=0$.

(For different modified gravity theories, the only difference is that this interaction term is characterized by the corresponding model dependent parameters. For example, in ECKS gravity $\mathcal{I} (=-4\varphi\rho_r)$ \cite{Kranas:2018jdc} is characterized by torsion scalar function $\varphi=\varphi(t)$ while fractal function $v=v(t)$ characterizes the interacting term $\left(-\dfrac{4}{3}\dfrac{\dot{v}}{v}\rho_r\right)$ \cite{Calcagni:2009kc} for fractal gravity.)

So the conclusion that the adiabatic thermodynamic prescription is not consistent with quasi-stable nature of radiation fluid in case of warm inflation, is independent of the choice of gravity theories. Now this problem can be solvable by changing the nature of the inflaton fluid.

One can note that this problem is basically involved with equation (\ref{eq10}). In the process of writing  equation (\ref{eq10}) from equation (\ref{eq9}), one must consider $\rho_I+p_I\neq0$. As non-equilibrium thermodynamics is considered, so $p^c_I\neq0$. Hence if one assume $\rho_I+p_I=0$, one cannot consider the adiabatic process in background of non-equilibrium thermodynamics.

Therefore, one can solve this problem by considering the inflaton fluid as variable cosmological constant and consequently $\rho_\phi+p_\phi=0$. So the Gibbs equation (\ref{eq9}) for inflaton fluid takes the  form
\begin{equation}
n_\phi T_\phi \dot{s}_\phi=-\Theta p^c_\phi
\end{equation}

Moreover, since the universe is in equilibrium state as a mixture of radiation and inflaton fluid, $\dot{s}_r\neq0$ and the Gibbs equation for radiation fluid can be written as
\begin{equation}
n_rT_r\dot{s}_r=-\Theta p^c_r-\frac{4}{3}\Pi_r\rho_r
\end{equation}

Using the expression for dissipative pressure for radiation fluid and quasi stable condition, the particle creation rate for radiation fluid can be written as
\begin{equation}
\Pi_r=3H-\frac{3Hn_rT_r\dot{s}_r}{\Gamma\dot{\phi}^2}
\end{equation} 

So the radiation particle number evolution is given by
\begin{equation}
\frac{\dot{n}_r}{n_r^2}=-\frac{3HT_r\dot{s}_r}{\Gamma\dot{\phi}^2}\neq0
\end{equation}

Hence radiation particle number is not conserved. In fact as $\dot{s}_r<0$, radiation particle is created. The particle number of radiation fluid can be written as
\begin{equation}
\frac{{n_r}_0}{n_r}=1+{n_r}_0\int \frac{3T_r\dot{s}_r}{a\Gamma\dot{\phi}^2}da
\end{equation}

Thus, it is reasonable to assume as a hypothesis that if radiation is produced in warm inflation exactly in the form of an adiabatic process, then a non-equilibrium thermodynamics analysis leads to a contradiction. However, this cannot happen in a warm inflation realization or equivalently, this does not hold in any realistic model of warm inflation. As far as the slow-roll conditions apply during warm inflation the adiabatic process is only at best, an approximation valid up to slow-roll coefficients. When the slow-roll approximation starts to be violated, the stronger will become the variations in temperature, radiation etc. as naturally happens at the end of inflation. Lastly, it worth to mention that when the slow-roll contributions are taken into account in the derivations, the above contradiction found is naturally resolved. Moreover, an alternative way to resolve this issue is to introduce a variable cosmological constant which accommodates the quasi-stable process in warm inflation with non-equilibrium thermodynamic description.

\textbf{Acknowledgement:}
The author A.B. acknowledges UGC-JRF (ID:1207/CSIRNETJUNE2019) and S.C. thanks Science and Engineering Research Board (SERB), India for awarding MATRICS Research Grant support (File No.MTR/2017/000407).
	           
\end{document}